\documentstyle[iopconf1]{article}

\newcommand{\mrm}[1]{\mbox{\rm #1}}
\newcommand{\beq}{\begin{equation}}
\newcommand{\eeq}{\end{equation}}

\newcommand{\bea}{\begin{eqnarray}}
\newcommand{\eea}{\end{eqnarray}}

\newcommand{\rfn}[1]{(\ref{#1})}

\newcommand{\np}[1]{Nucl. Phys. {\bf #1}}
\newcommand{\pl}[1]{Phys. Lett. {\bf #1}}
\newcommand{\pr}[1]{Phys. Rev. {\bf #1}}
\newcommand{\prl}[1]{Phys. Rev. Lett. {\bf #1}}

\newcommand{\sm}{SM}

\newcommand{\xs}{cross section}

\newcommand{\pe}{\mbox{$e^+e^-$}}
\newcommand{\ee}{\mbox{$e^-e^-$}}
\newcommand{\mm}{\mbox{$\mu^-\mu^-$}}
\newcommand{\ep}{\mbox{$e^-\gamma$}}

\newcommand{\bil}{bilepton}

\newcommand{\guts}{grand unified theories}

\def\lsim{\mathrel{\vcenter{\hbox{$<$}\nointerlineskip\hbox{$\sim$}}}}
\def\gsim{\mathrel{\vcenter{\hbox{$>$}\nointerlineskip\hbox{$\sim$}}}}

\begin{document}

\hfill FTUV/97-55

\hfill IFIC/97-57

\hfill hep-ph/9709204

\title{Bileptons - Present Constraints and Future Prospects at Colliders}

\author{ Martti Raidal\footnote{E-mail: raidal@titan.ific.uv.es}}

\affil{Department of Theoretical Physics, University of Valencia, 
46100 Burjassot, Valencia, Spain}

\beginabstract
Bosons which couple to two leptons and carry lepton number $L=2,$
generically called bileptons, appear in many extensions of the 
\sm. We review the present constraints on bilepton 
parameters coming from low and high energy data. In particular,
we point out that the amplitude of $\mu-e$ conversion in nuclei is 
enhanced by large logarithms if compared with the $\mu\to e\gamma$
one. Bilepton couplings will be tested most stringently at future 
colliders. For wide range of neutrino masses the cross section 
of one of the processes $e^-e^- (\mu^-\mu^-)\to \ell_i^-\ell_i^-,$ 
$i=e,\mu, \tau,$ mediated by scalar triplet bilepton, has such a
 {\it lower} bound that it should be observed at future facilities.
\endabstract

\section{Introduction}

One of the peculiar features of the standard model (\sm)  
is that none of its bosons 
carry global quantum numbers. This is no longer the case in
 wide range of  extensions of the \sm,
such as \guts~\cite{guts},
theories with enlarged Higgs sectors~\cite{higgs},
theories which generate neutrino Majorana masses~\cite{maj}
as well as technicolour theories~\cite{tc} and 
theories of compositeness~\cite{comp}. 
Following Ref. \cite{cd} \bil s are defined to be bosons 
which couple to two leptons (but not to quarks) 
and which carry two units of lepton number.
They can be both scalars and vectors.
Their interactions need not necessarily conserve lepton flavour,
but otherwise  the symmetries of the \sm\ should be respected.
The most general $SU(2)_L\times U(1)_Y$ invariant 
renormalizable dimension four Lagrangian  for bileptons 
is given by~\cite{cd}
\bea
\label{lag}
{\cal L}&~=~&-~
\lambda_1^{ij} \quad L_1^{+} \quad 
\left( \bar \ell^c_i P_L \nu_j - \bar \ell^c_j P_L \nu_i \right) \\\nonumber
&&+~
\tilde\lambda_1^{{ij}} \quad \tilde L_1^{++} 
\quad \bar \ell^c_i P_R \ell_j \\\nonumber
&&+~
\lambda_2^{ij} \quad L_{2\mu}^{+} 
\quad \bar\nu^c_i \gamma^\mu P_R \ell_j \\\nonumber
&&+~
\lambda_2^{ij} \quad L_{2\mu}^{++} 
\quad \bar \ell^c_i \gamma^\mu P_R \ell_j \\\nonumber
&&+~
\lambda_3^{{ij}} \quad L_{3}^{0} 
\quad \bar\nu^c_i P_L \nu_j \\\nonumber
&&-~
\lambda_3^{{ij}}/\sqrt{2} \quad L_{3}^{+} 
\quad \left( \bar \ell^c_i P_L \nu_j + \bar \ell^c_j P_L \nu_i \right) 
\\\nonumber
&&-~
\lambda_3^{{ij}} \quad L_{3}^{++} 
\quad \bar \ell^c_i P_L \ell_j \\\nonumber
&&+~
\mbox{ h.c.}~ ~,
\eea
where the subscripts 1--3 label the dimension
of the $SU(2)_L$ representation
which the \bil s belong to
and the indices $i,j=e,\mu,\tau$ stand for the lepton flavours.
In this talk I assume the couplins $\lambda^{ij}$ to be real. In that
case $\lambda$ is anti-symmetric for scalar singlets, symmetric for
scalar triplets and arbitrary for vector \bil s. 

In the following I shall concentrate on the doubly charged components
of \bil\ multiplets. 
First I shall review the present most stringent bounds on
their parameters and later discuss their production at future colliders.

\section{Present constraints}

\begin{table}
\begin{center}
\caption{Upper bounds on diagonal \bil\ couplings for $m_B=1$ TeV.}
\begin{tabular}{ccc}
\topline
 & Combination &  \\
Process & of couplings & Upper bound \\
\midline
M{\o}ller & $\lambda^{ee}$ & 4 (2) \\
$(g-2)_\mu$ & $\lambda^{\mu \mu}$ & 10 (5) \\
$M-\bar{M}$ & $\lambda^{ee}\lambda^{\mu \mu}$ & 0.2 (0.2) \\
\bottomline
\end{tabular}
\end{center}
\end{table}
Due to large bilepton masses, $m_B \gsim \cal{O}(100)$ GeV 
(LEP II pair production threshold), and unknown couplings to 
leptons one can presently only constrain \bil\ effective  couplings
of a generic form  $G=\sqrt{2}\lambda^2/(8m^2_B).$  
Negative results in searches for 
 the lepton flavour violating processes 
$\ell_l\rightarrow 3\ell_f$ and $\ell_l\rightarrow \gamma \ell_f,$
where $l=\mu, \; \tau,$ and $f=e,\; \mu,$ put orders of magnitude
more stringent bounds on  off-diagonal bilepton couplings than 
one obtains from M{\o}ller scattering, $(g-2)_\mu$ studies as well as
from the searches for muonium-antimuonium conversion for diagonal
couplings $\lambda^{ee}$ and $\lambda^{\mu \mu}.$
To date there is no constraints on  
$\lambda^{\tau\tau}$ without involving off-diagonal elements.

Most stringent  constraints on diagonal \bil\ couplings  \cite{cd,willy,lnv}
are summarized in Table 1. 
The limits are approximately equal for singlet and triplet scalars
and given as the first numbers in the table  while the limits for vector
\bil s are presented in brackets.
Here and in the following \bil\ masses are always taken in units of 1 TeV. 
As can be seen, for the scale of new physics  $\cal{O}(1)$ TeV
only the muonium-antimuonium conversion experiment \cite{willy} constrains
diagonal \bil\ couplings in a meaningful way.

\begin{table}
\begin{center}
\caption{Upper bounds on \bil\ couplings for $m_B=1$ TeV from various low
energy leptonic processes.}
\begin{tabular}{ccc}
\topline
 & Combination &  \\
Process & of couplings & Upper bound \\
\midline
$\mu\to 3e$ & $\lambda^{\mu e}\lambda^{ee}$ & $4 (2)\cdot10^{-5}$ \\
$\mu\to e\gamma$ & $(\lambda\lambda)^{\mu e}$ & $1.2 (1.2) \cdot10^{-2}$ \\
$\mu-e$ & $(\lambda\lambda)^{\mu e}$ & $7 (7)\cdot10^{-3}$ \\
\bottomline
\end{tabular}
\end{center}
\end{table}
Most stringent constraints on 
combinations of couplings $\lambda$ from other lepton
flavour violating processes \cite{cd,lnv,san} are summarized in Table 2. 
For shortness we have presented the processes involving first two 
generations only.
Constraints from the processes involving third  generation are considerably
weaker. 
As expected, the strongest limit derives from $\mu\to 3e$ since it can 
occur at tree level. However, it constrains only a particular combination of
couplings. On the other hand, 
$\mu\to e\gamma$ and $\mu-e$ conversion limits apply
on a sum of coupling constant products since all possible 
intermediate states can occur at loop level and, therefore,
provide complementary information. 

One should note here that contrary to the claim in Ref.\cite{ver}
$\mu-e$ conversion is more sensitive to \bil\ interactions than 
$\mu\to e\gamma.$  The general 
$\bar{e}\mu\gamma$ vertex can be parametrized in terms of form factors as 
follows
\beq
j^\rho=\bar{e}\left[ \gamma_\mu (f_{E0}+f_{M0}\gamma_5)
\left( g^{\mu\rho}- \frac{q^{\mu}q^{\rho}}{q^2}\right) + 
(f_{M1}+f_{E1}\gamma_5) i\sigma^{\rho\nu}\frac{q_{\nu}}{m_{\mu}}\right]\mu,
\eeq
where $q$ is the  momentum transferred by photon. 
While $\mu\to e\gamma$ is induced only by form factors proportional to
$\sigma^{\rho\nu}$ term then $\mu-e$ conversion rate is proportional to
$(|f_{E0}+f_{M1}|^2+|f_{M0}+f_{E1}|^2).$ Since the form factors
$f_{E0}$ and $f_{M0}$ are enhanced by large 
$\ln(m_{\ell}^2/m^2_B)\sim \cal{O}(10),$ 
where $m_{\ell}$ stands for the charged lepton mass
running in the loop,  if compared with $f_{E1}$ and $f_{M1}$ then
$\mu-e$ conversion is enhanced  while $\mu\to e\gamma$ is not. 
The logarithmic factors arise from the diagrams in which photon is 
attached to the charged lepton. Therefore, one expects logarithmic
enhancement of $\mu-e$ conversion also in models with, e.g., 
leptoquarks  or broken $R$-parity  
but not in models considered in
Ref.\cite{ver}.

\section{Production at colliders}

The most stringent constraints on \bil\ couplings and masses will be 
obtained at future linear and muon colliders \cite{gun10}. 
Bileptons with masses
$m_B\lsim \sqrt{s}/2$ can be pair produced in \pe\ collisions almost
model-independently due to their couplings to photon \cite{gun11}. 
Single production
of \bil s in \pe\ and \ep\ collisions allowes one to probe masses 
up to $\sqrt{s}$ and test all the couplings $\lambda^{ij}$ more 
sensitively than any of the present low energy experiments \cite{mina1}.
The most appropriate  for studying \bil s are
\ee\ and \mm\ collision modes in which resonant s-channel production
of \bil s via the process 
\beq
\ee (\mm)\to \ell_i^-\ell_i^-,
\label{proc}
\eeq
 $i=e,$ $\mu,$ $\tau,$
is possible \cite{ee2bil,jack}.  For realistic machine and beam
parameters sensitivity of $\lambda^{ij}\lsim 5\cdot 10^{-5}$
can be acheved \cite{jack}. Vector \bil s can be discriminated from
the scalar ones by studying photon distribution in $\ee\to L^{--}\gamma.$
 
Despite of this extraordinary sensitivity it could happend that due to  
small $\lambda$'s or high bilepton  masses no positive signal 
will be detected at future colliders.
However, this may not be the case if neutrinos are massive as
predicted by most of the extensions of the \sm\ and the process 
\rfn{proc} is mediated by triplet scalar \bil.

 If the
sum of light  neutrino masses exceeds $\sim 90$ eV at least one of 
them has to be unstable. In order not to overclose the Universe
the lifetime and mass of such an unstable neutrino $\nu_l$
must satisfy the requirement
\cite{eight} 
\bea
\tau_{\nu_l} \lsim 8.2 \cdot 10^{31} \; \mbox{MeV}^{-1} \left(
\frac{100 \; \mbox{keV}}{m_{\nu_l}} \right)^2 .
\label{constr}
\eea
The radiative decay modes 
$\nu_l \to  \nu_f \gamma, \; \nu_f\gamma\gamma $ 
are highly suppressed \cite{rad}
and cannot satisfy  the constraint (\ref{constr}). The same is also 
true for $Z'$ contribution to $\nu_l \to 3 \nu_f$ decay \cite{zprim}.
The only possibility which lefts over is  the decay
$ \nu_l   \to   3 \nu_f $ due to neutrino mixings
induced by the neutral component of triplet \bil.

Clearly, since $L^0$ and $L^{--}$ belong to the same multiplet the 
reaction \rfn{proc} can be related to decay $ \nu_l   \to   3 \nu_f.$ 
Since the latter decay rate is bounded from below by the 
constraint \rfn{constr} and limits on neutrino mixings also 
the \xs\ of \rfn{proc} has a lower limit. 
Let us consider numerically the most conservative case,
$ \nu_\tau   \to   3 \nu_e.$
From the constraints presented  above one  obtains  
the following bound on $G^{e\tau}$ which induces
the process $e^-e^-\to \tau^-\tau^-$ \cite{mina2}
\beq
G^{e\tau}\gsim 2\cdot 10^{-3} \frac{\lambda_{\tau\tau}}
{\lambda_{ee}+\lambda_{\tau\tau}} 
\;\mrm{TeV}^{-2}\;,
\label{gpminte}
\eeq  
where $G^{e\tau}=\sqrt{2}\lambda^{ee}\lambda^{\tau\tau}/(8m^2_B).$
On the other hand, studies of the process \rfn{proc} at linear colliders give
that far off resonance the minimal testable $G^{ij}$ are
$G^{ij}(\mrm{min})=1.4\cdot 10^{-4}/s 
\;\mrm{TeV}^{-2}\;, $
approximately the same for all relevant $i,j.$
Therefore, the process $e^-e^-\to \tau^-\tau^-$ should be
detected at the 1 TeV linear collider unless 
$\lambda_{\tau\tau}/\lambda_{ee}\lsim  10^{-1}.$
On the other hand, if this is the case then
\beq
G^{ee}\gsim 2\cdot 10^{-3}  
\;\mrm{TeV}^{-2}\;,
\label{gpminee2}
\eeq  
and  the  excess of the electron 
pairs due to the s-channel bilepton exchange will be detected. 
Note that the positive signal should be seen if $\sqrt{s}\gsim 0.3$ TeV
which is below the planned initial energy of the linear collider.

Similar argumentation applies to all possible neutrino decays 
with even higher lower limits. Therefore, if one of the neutrino mass,
indeed, exceeds 90 eV one should detect at least one of the
processes \rfn{proc} at future colliders.

\section{Conclusions}

Bileptons appear in many realistic extensions of the \sm.
The present constraints on diagonal \bil\ couplings are
rather weak compared to the bounds on off-diagonal couplings. 
Some of the latter ones, $(\lambda\lambda)^{e\mu},$
 will be tested about two orders of magnitude more 
stringently in planned $\mu-e$ conversion experiments.
This process is most sensitive to the couplings due to logarithmic 
enhancement of the conversion amplitude. 

Future linear and muon colliders will be more sensitive to \bil\
couplings than any of the present experiments. In particular,
if triplet scalar particles exist and some neutrino masses exceed 90 eV 
then the doubly charged component of the  triplet should be
detected in \ee\ or \mm\ collision modes of the colliders.

\section*{Acknowledgments}
I would like to thank Organizers for creating pleasant atmosphere at 
the Ringberg castle. I have benefitted from discussions with P. Herzceg
and especially with F. Cuypers and A. Santamaria whom I thank warmly.

\end{document}